\documentclass[fleqn,12pt,twoside]{article}
 \usepackage{espcrc1}
\usepackage{graphicx}% Include figure files
\usepackage{epsf,epsfig,latexsym}

\usepackage{graphicx}
\title{Studying QCD at low $x$ and high mass with CMS}
\author{Michael Murray, University of Kansas, for the CMS Collaboration}

\begin{document}
\maketitle

\begin{abstract}
The CMS heavy ion program can study quark matter over an unprecedented range of Bjorken $x$ and mass. CMS is equipped with excellent  detectors to exploit the new physics probes available at $\sqrt{s_{NN}}=5.5$TeV. The high rate capability and wide rapidity coverage will allow us to study the 
correlations between low $x$ and high mass phenomena. 
\end{abstract}

\begin{figure}[h]
\begin{center}
\begin{minipage}[h]{75mm}
\epsfig{file=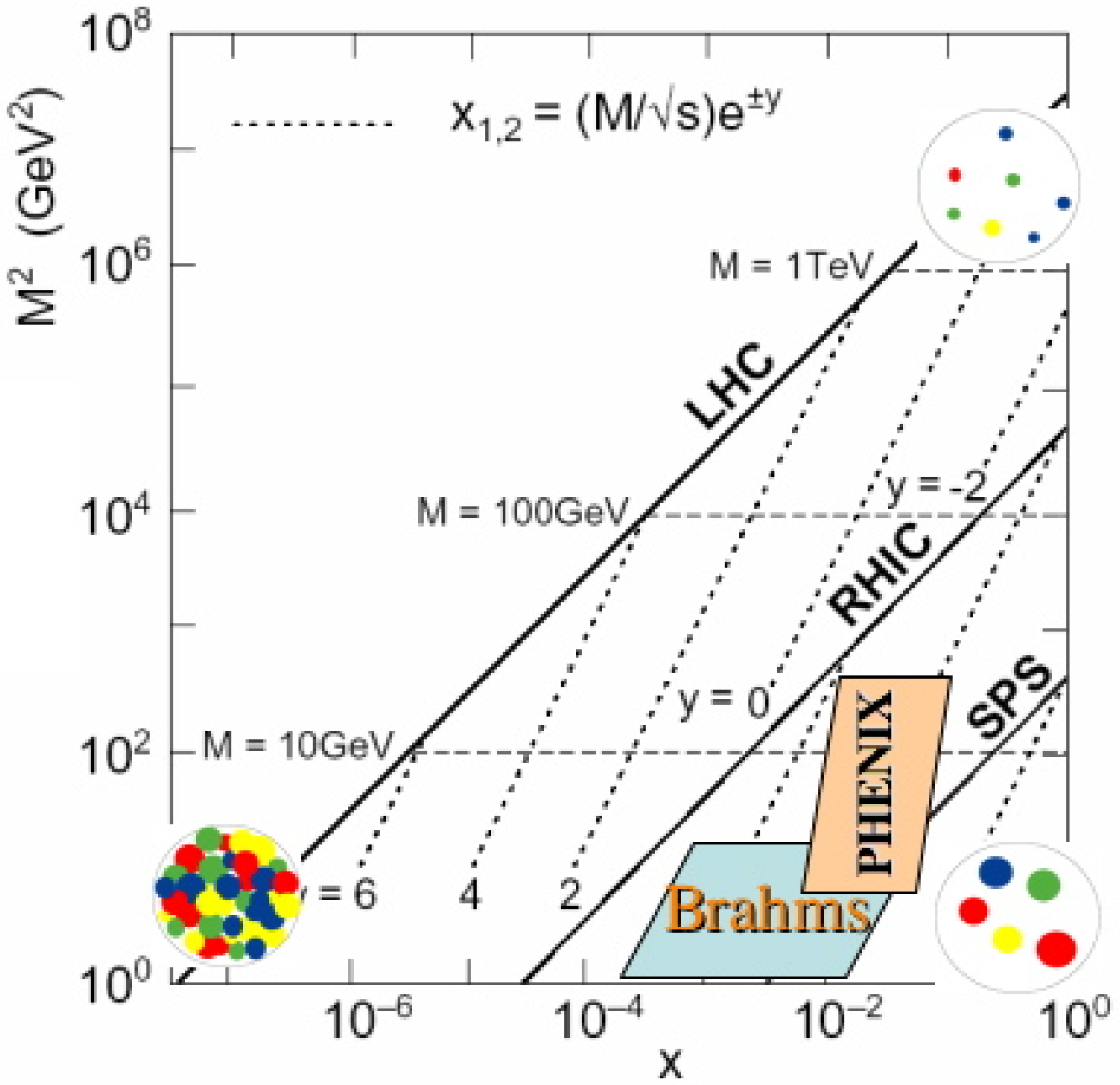,width=75mm}
\caption{\label{XvMass} The kinematic reach of the LHC.
% with schematic representations of the gluon density in a nucleus. 
The forward CMS detectors cover  $\eta \le 7$. The powerful trigger and high rate capability allow us to study very high mass objects.
}
\end{minipage}
\hspace{\fill}
\begin{minipage}[h]{75mm}
\epsfig{file=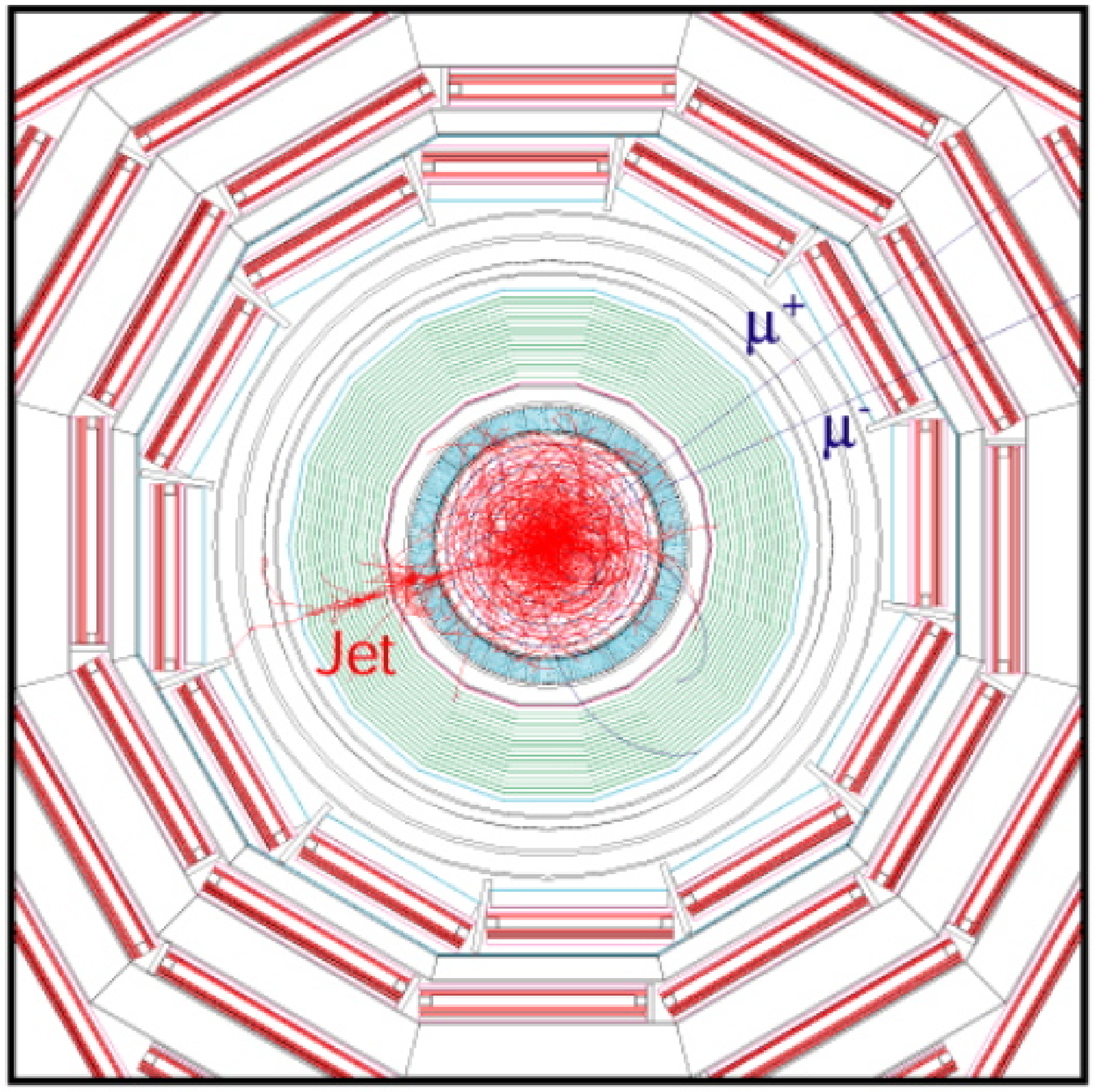,width=75mm}
\caption{\label{Z02muon}
 Reconstruction in CMS of a central PbPb event with a 
 Z$^0 \rightarrow \mu^+\mu^-$ opposite a jet.}
\end{minipage}
\end{center}
\end{figure}
\vspace{-1.0cm}

The Large Hadron Collider (LHC) will collide lead nuclei at %$\sqrt{s_{NN}}=5.5$ TeV, an 
energies 27 times larger than RHIC. It seems probable  that this increase in energy  
 will produce matter that is significantly different from the opaque ideal fluid discovered at RHIC \cite{WhitePapers}. %We expect that 
The LHC experiments should reveal the long sought for weakly interacting quark gluon plasma; another manifestation of partonic matter, 
different from the strongly interacting QGP 
produced at RHIC. The increase in energy, coupled with the very wide 
%(pseudo)-rapidity 
angular range will allow CMS to probe the initial state of the %colliding 
nuclei down to  very low Bjorken $x$. In this region, 
 the gluons are expected to form a classical field, 
dubbed the ``Color Glass Condensate"  \cite{MclVenu,BrRdA}. 
This state may represent the high energy limit of QCD.

Figure \ref{XvMass} shows the increase in phase space of the LHC compared to the RHIC and SPS accelerators. CMS can study heavy ion collisions to near the limits of this space.
CMS also does an excellent job of reconstructing hard
probes such as jets and $Z^0$s in $pp$, $pA$ and $AA$ collisions, see
Fig.~\ref{Z02muon}.  Thus CMS can study QCD over a wide range of $x$ and mass.  %, $M$.
%The increased energy of the LHC will make available abundant, precise and %new probes of the created matter Ð far beyond those which can be studied %at RHIC or its planned descendents. 
CMS will provide differential measurements allowing for detailed quantitative tests of theory including:
\begin{itemize}
\item The study of fundamental QCD processes and parton densities over the widest available $x$ range. This study will begin with the calibrated region at midrapidity and move to uncharted territory at forward angles to search for novel features of the initial partonic state such as the Color Glass Condensate.
\item Flavor tagged jets over a wide $p_T$ range since their correlations and properties will be decisive in testing our understanding of parton energy loss process. 
\item Formation and decay of heavy quarkonia, sensitive to the specific properties of the medium.
\item Day One measurements of the statistical and thermal properties of the created matter including  energy and particle flow and  equilibration.  
\end{itemize}

\begin{figure}%[t]
\epsfig{file=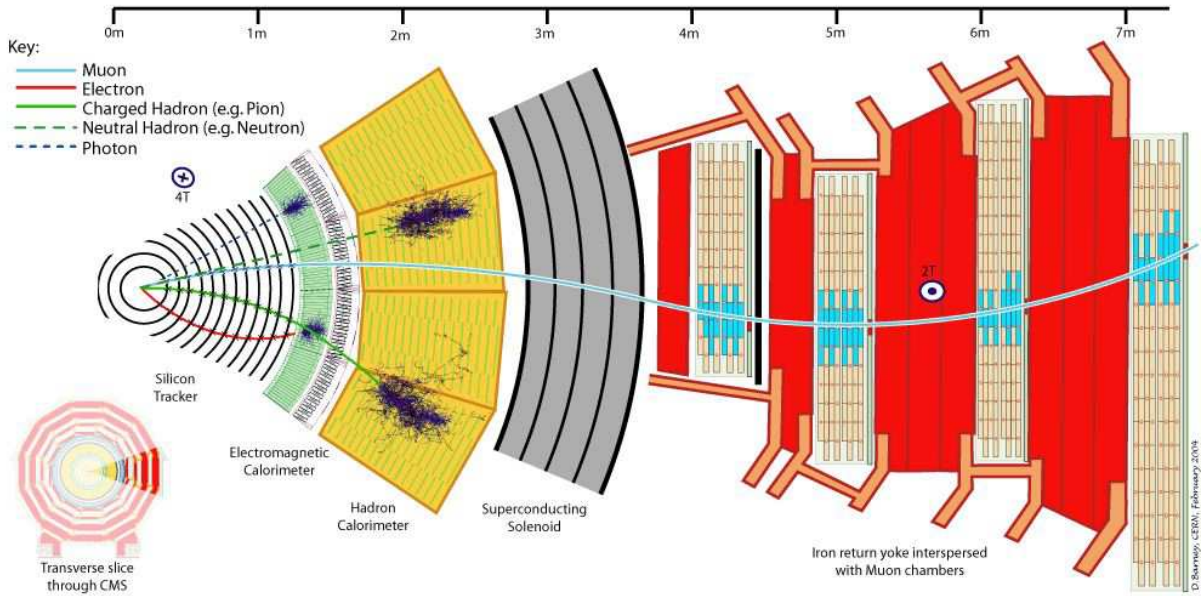,width=\columnwidth}
\caption{\label{CmsCore} The central CMS detectors.  The strengths of the experiment are in high resolution tracking, fine grained electromagnetic calorimetry, 4 T magnetic field and very high momentum resolution for muons.
}
\end{figure}

Figure \ref{CmsCore} shows the central detectors of CMS \cite{CmsPub}. These are designed to disentangle up to 40 $pp$ collisions of $\sqrt{s} = 14$ TeV (7+7) each in a single bunch crossing. The inner (pixel) and outer (strip) trackers as well as the electromagnetic and hadronic calorimeters are all inside the 4 T magnetic field. This allows for excellent momentum resolution. The return flux of the magnet is exploited to deflect the muons into an ``S" shape to allow an independent momentum measurement. 

The innermost tracker is  composed of 3 barrels  and 2 endcaps of pixels. 
The 3 barrels cover  $|\eta| \le 2.1$ 
%(radial distances from beam line : 4,7, 11 cm) 
and  are made of more than 9.6, 16 and 22.4 million 
pixels respectively.
For central Pb+Pb collisions the  pixel occupancy is not 
expected to exceed $2\%$ at the innermost radius of 4cm. 
 The position resolution of high $p_T$ tracks is 15-20~$\mu$m.
The inner silicon strip tracker counter consists of 4 cylindrical layers in the barrel 
and 3 disks in each endcap. The first two layers in both the barrel and endcap are double sided while the rest are single sided. The outer tracker consists of 6 layers in the barrel and 9 in the endcap. Again the first two layers in both regions are double sided. In addition layer 5 of the endcap is also double sided.  
%For the barrel layers 1,2,5 and 6 are double sided and the rest single %sided. In the endcap layers 1, 2 and 5 are double sided. 
%The outer tracker has 6 layers of silicon detectors in the barrel region
%and 9 disks in the endcap.
 At the nominal field of 4 T,  particles with
$p_T < 800$ MeV curl up in the tracker. 
 Above $p_T = 1$ GeV we expect to reconstruct 
$~80-90\%$ of the tracks in central Pb+Pb events, 
almost ndependent of $p_T$. 
%The  resolution is $<2\%$ 
For $|\eta| < 0.7$ $\sigma p_T < 2\%$
%The  resolution is $<2\%$ for $|\eta| < 0.7$. 
%This efficiency is almost independent of $p_T$. 

The electromagnetic calorimeter is made of  83000  
crystals of PbWO$_4$ scintillator each 25.8 radiation lengths deep. The barrel
 covers $|\eta|<1.5$  with 
$\Delta\eta \times \Delta\phi$~=~0.0175$\times$0.0175. 
The granularity gradually coarsens in
the endcaps, reaching 
$\Delta\eta \times \Delta\phi$~=~0.05$\times$0.05 at $|\eta|=3.0$. 
The endcap also has a silicon preshower detector in front of it. 
The hadron calorimeter ($|\eta| \le 3.0$) is made of scintillator and copper plates and is about  5.2 nuclear 
interaction lengths deep at $\eta=0.$ 
A pair of   forward calorimeters HF,  $\pm$11~meters from the interaction point, cover $3 \le |\eta| \le 5$. HF uses steel as an absorber and Cerenkov fibers as the active medium. The calorimeter has  depth of  
$~9.5$ interaction lengths with a front electromagnetic section 15 radiation lengths deep.

%There are at least 10$\lambda$ of calorimeters before the first station and %an additional 10$\lambda$ of iron yoke before the last station.

\begin{figure}[t]
\begin{center}
\begin{minipage}[h]{65mm}
\epsfig{file=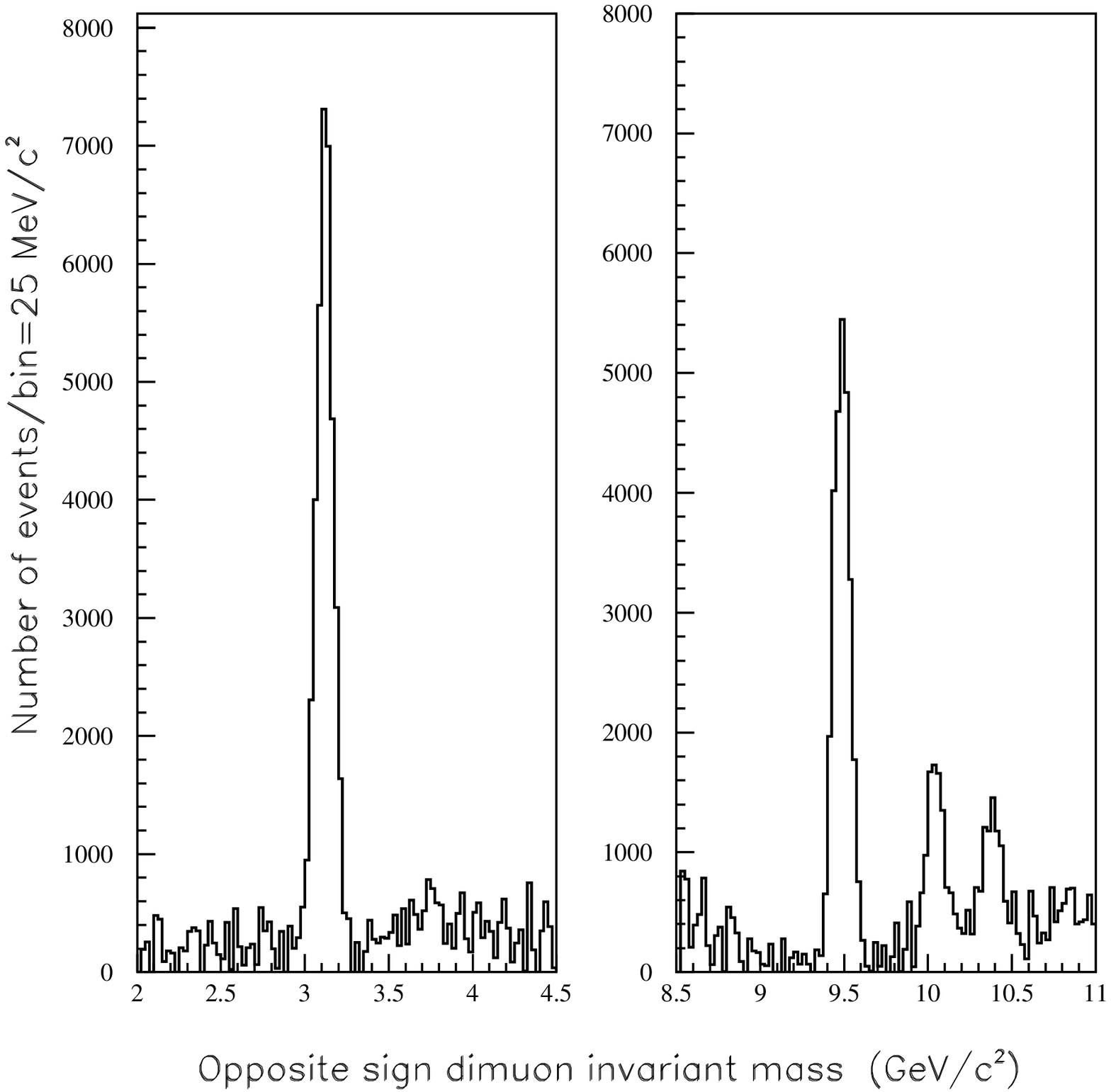,width=65mm}
\caption{\label{DiMuonMass}
Quarkonia yields as a function of the muon invariant mass. The like-sign background has been subtracted.}
\end{minipage}
\hspace{\fill}
\begin{minipage}[h]{80mm}
\epsfig{file=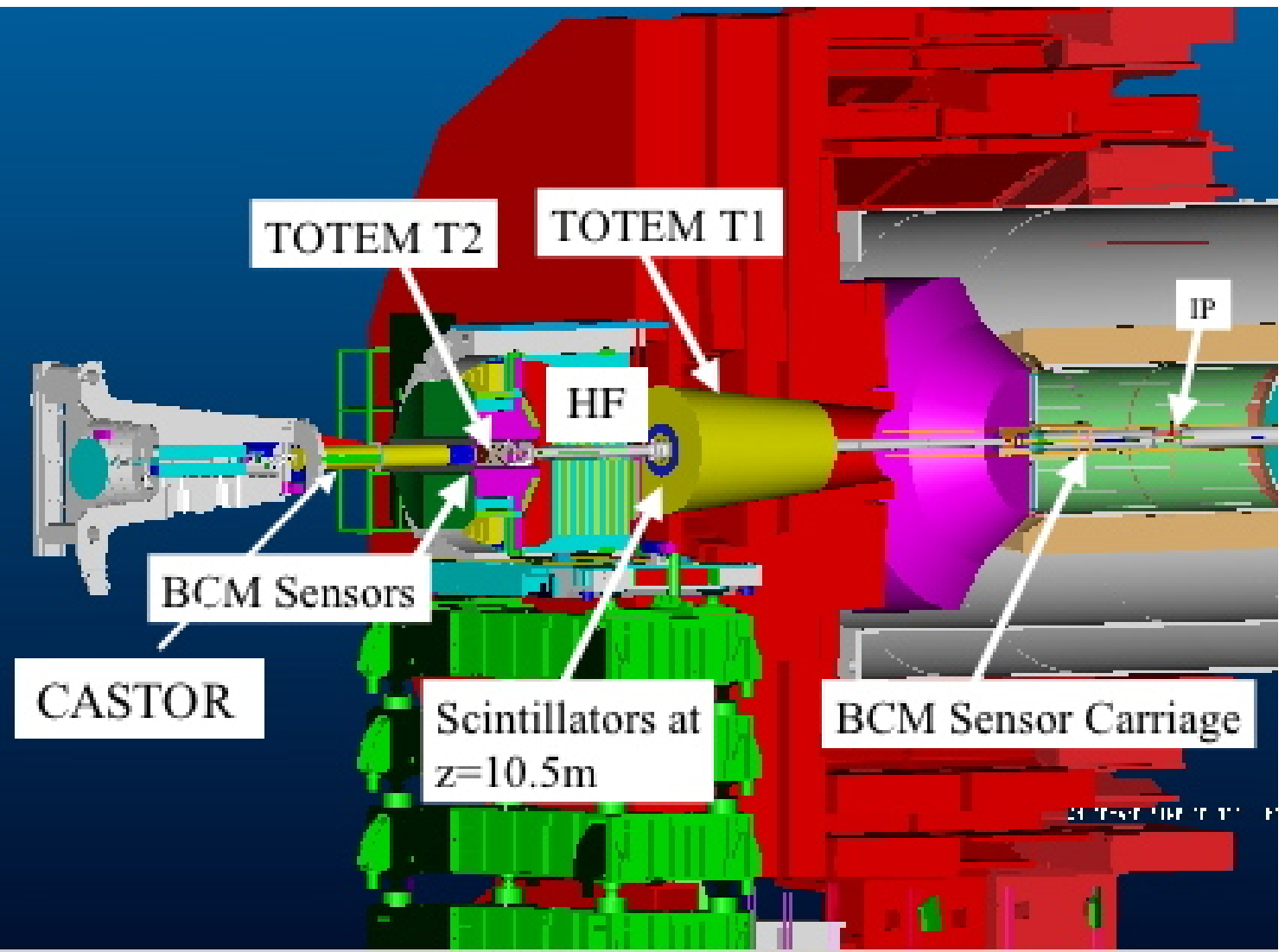,width=80mm}
\caption{\label{CmsForward}
The forward detectors in CMS.
The interaction point, IP, is at the right, followed by the Beam Condition Monitors, BCM, and the T1 and T2 multiplicity detectors.
%The forward calorimeter, 
HF, measures the energy 
of the particles that hit T1 while CASTOR does the same for T2. The ZDC sits 140 m downstream.
}
\end{minipage}
\end{center}
\end{figure}

The muon system uses the inner detectors as a shield to absorb hadrons, photons and electrons. Additional shielding comes from the iron yoke plates that provide the flux return for the magnet. This returning magnetic flux bends the 
muons in the opposite direction that they had in the inner detector. Four tracking stations are arranged in rings in the barrel and as disks in the 
endcaps.  They are shown in orange in Figs.~\ref{Z02muon} and~\ref{CmsCore}.  
The muon system can reconstruct $\Upsilon$, $\Upsilon'$ $\Upsilon''$ 
in Pb+Pb collisions with a mass resolution of about 50 MeV 
as shown in Fig.~\ref{DiMuonMass}.

Figure \ref{CmsForward} shows the forward detectors. These can make an almost complete measurements of the flow of charged particles and energy into the forward region. CASTOR ($5 \le \eta \le 7$) \cite{CASTOR} and the 
ZDC $(\eta \ge 8.7)$ are two Cerenkov tungsten calorimeters  while T1 and T2 are multiplicity counters. T2 should be useful for heavy ion running since it 
is a GEM detector with high granularity and analog readout \cite{TOTEM}. 

%\vspace{-3.0cm}
A particular strength of CMS is the flexibility of its trigger/data acquisition system, designed to handle 14 TeV $pp$ collisions at a luminosity  
$ L=10^{34}{\rm cm^{-2} s^{-1}}$. 
CMS uses commodity network switches to combine data from its subdetectors and large farms of commodity processors to select physics events. This  design allows the elimination of the traditional ``Level-2" trigger that requires custom processors and inflexible software. By concentrating the event selection tasks in a relatively simple ``Level-1" followed by a powerful ``High Level Trigger", the experiment achieves an unprecedented level of flexibility and selection control.
The multiplicity per event is much higher for Pb+Pb runs but 
the luminosity is much lower $L=10^{27}{\rm cm^{-2} s^{-1}}$.  
Thus the high level trigger should be able to reconstruct almost all heavy ion events in real time.

The CMS heavy ion program thus has unique capabilities for heavy ion physics. 
Excellent tracking and calorimetry are located within a strong magnetic 
field for precise momentum measurements. The muon system gives an unprecedented mass resolution for quarkonia and $Z^0$s. The fast and flexible trigger 
helps optimize the available luminosity. 
 The very wide rapidity coverage is essential to study the initial Color Glass Condensate while hard probes such as $\Upsilon$ and jets should peer deep into the quark gluon plasma.  

\end{document}